\documentclass[11pt]{article}  
\usepackage{amssymb,graphicx} 
\usepackage {apalike}
\usepackage{undertilde}
\usepackage[normalem]{ulem} 
\usepackage{amscd}
\usepackage {amsfonts}
\usepackage{cancel}
\begin{document}

\title{Einstein's Equations for Spin $2$ Mass $0$ from Noether's Converse Hilbertian Assertion} %

\maketitle

\begin{center}
 \author{J. Brian Pitts \\ Faculty of Philosophy, University of Cambridge 
\\ jbp25@cam.ac.uk \\
forthcoming in \emph{Studies in History and Philosophy of Modern Physics} }
\end{center}


\begin{abstract}

An overlap between the general relativist and particle physicist views of Einstein gravity is uncovered.  Noether's 1918 paper developed Hilbert's and Klein's reflections on the conservation laws.  Energy-momentum is just a term proportional to the field equations and a ``curl'' term with identically zero divergence.  Noether proved a \emph{converse} ``Hilbertian assertion'':  such ``improper'' conservation laws imply a generally covariant action. 
 
Later and independently, particle physicists derived the nonlinear Einstein equations assuming the absence of negative-energy degrees of freedom (``ghosts'') for stability,  along with universal coupling:  all energy-momentum including gravity's serves as a source for gravity.  Those assumptions (all but) imply (for 0 graviton mass) that the energy-momentum is only a term proportional to the field equations and a symmetric curl, which implies the coalescence of the flat background geometry and the gravitational potential into an effective curved geometry.  The flat metric, though useful in Rosenfeld's stress-energy definition, disappears from the field equations.  Thus the particle physics derivation uses a reinvented Noetherian converse Hilbertian assertion in Rosenfeld-tinged form.  

The Rosenfeld stress-energy is identically the canonical stress-energy plus a Belinfante curl and terms proportional to the field equations, so the flat metric is only a convenient mathematical trick without ontological commitment.  Neither generalized relativity of motion, nor the identity of gravity and inertia, nor substantive general covariance is assumed.  The more compelling criterion of lacking ghosts yields substantive general covariance as an output.  Hence the particle physics derivation, though logically impressive, is neither as novel nor as ontologically laden  as it has seemed.  

\end{abstract}  %

keywords: conservation laws; General Relativity; Noether's theorems; energy-momentum tensor; particle physics; Belinfante-Rosenfeld equivalence



\section{Introduction}

It is often held that there is a great gulf fixed between the views of gravitation held by general relativists and those held by particle physicists.  Carlo Rovelli has commented on  the resultant effects on quantum gravity research programs. \begin{quote} 
The divide is particularly macroscopic between the covariant line of research
on the one hand and the canonical/sum over histories on the other. This divide
has remained through over 70 years of research in quantum gravity. The
separation cannot be stronger.\ldots 
Partially, the divide reflects the different understanding of the world that the
particle physics community on the one hand and the relativity community on the other hand, have. The two communities have made repeated and sincere
efforts to talk to each other and understanding each other. But the divide
remains, and, with the divide, the feeling, on both sides, that the other side is
incapable of appreciating something basic and essential\ldots. Both sides expect that the
point of the other will turn out, at the end of the day, to be not very relevant.\ldots 
 Hopefully, the recent successes of both lines will force the two sides, finally, to face the problems that the other side considers
prioritary\ldots \cite{Rovelli}.  \end{quote}

The following anecdote gives some background about the history of (non-)interaction between the general relativity and particle physics communities: \vspace{-.09in}
\begin{quote}
The advent of supergravity [footnote suppressed]  made relativists and particle physicists meet.  For many this was quite a new experience since very different languages were used in the two communities.  Only Stanley Deser was part of both camps.  The particle physicists had been brought up to consider perturbation series while relativists usually ignored such issues.  They knew all about geometry instead, a subject particle physicists knew very little about. 
\cite[p. 40] {BrinkDeserSupergravity}
\end{quote}
Unfortunately the pocket of convergence between the two communities in physics, though by now expanded well beyond Deser (and perhaps allowing for hyperbole), is still rather small.

If one can show that the gulf is not as large as it seemed, might it be easier to imagine quantum gravity programs that also split the difference?  This paper will not aim to say much about quantum gravity, but it will show that a certain part of the difference between general relativists' and particle physicists' views is illusory, because a key part of particle physicists' 1939-73 ``spin $2$'' derivation of Einstein's equations from flat space-time is basically Noether's 1918 converse Hilbertian assertion.  Hilbert and later Klein had found that the conserved energy-momentum in General Relativity consists of a term proportional to the Einstein tensor and hence having a value of $0$ using the Euler-Lagrange equations (vanishing on-shell, one says) and a term with automatically  vanishing divergence (a ``curl'').  They worked with the most straightforwardly derivable expressions for gravitational energy-momentum:  pseudo-tensors and what one now calls the Noether operator \cite{Hilbert1,KleinGREnergy1917,KleinGREnergy1918,PaisEinstein,OlverLie,RoweGoettingenNoether,RoweEinsteinBianchiIdentities,BradingBrownSymmetries}.  Noether also proved a converse to the ``Hilbertian assertion,'' along with converses to the two theorems associated with her name \cite{Noether}.

  Noether formulated the Hilbertian assertion and its converse as follows: 
\begin{quote} 
 If [action] $I$ admits of the displacement group, then the energy relationships
become improper if and {\bf only if} $I$ is invariant with respect to an infinite group containing
the displacement group as subgroup. [footnote suppressed]  \cite[emphasis added]{Noether}  \end{quote} 
The ``only if'' clause, the converse to which I call attention, offers the possibility of arriving at general covariance from improper conservation laws.  Such conservation laws consist only of terms vanishing by the field equations and terms with identically vanishing divergence (``curls'')---just the kinds of things that separate the metrical and canonical stress-energy tensors (Belinfante-Rosenfeld  equivalence), and just the kinds of things that one is often said to be free to modify to taste. Hence one could modify the stress-energy for vacuum General Relativity to be identically $0$ by that reasoning if one wished.
One need not agree that there is anything improper about `improper' conservation laws \cite{EnergyGravity} to use the familiar term.

During the 1920s-50s, the luster of General Relativity faded among physicists  \cite{EisenstaedtEtiage,Eisenstaedt,vanDongenBook,SchutzGR}, even if the popular image of Einstein was undiminished.  General Relativity was mathematical, hardly empirical,  and tied to protracted fruitless speculation about unified field theories.  Relativistic quantum mechanics and (at times) field theory were flourishing, by contrast.  In such a context it made sense, if one was not ignoring General Relativity, to try to subsume it into the flourishing framework of particle physics.  Apart from questions of quantization, this project succeeded.  In 1939 Fierz and Pauli noticed that the linearized vacuum Einstein equations were just the equations for a particle/field of spin $2$ and mass $0$ \cite{PauliFierz,FierzPauli,Fierz2}. Even Nathan Rosen, long-time collaborator with Einstein, appeared to be defecting in 1939:  he proposed to reduce the conceptual distance between General Relativity and other field theories by introducing a flat background geometry and wondered aloud whether the nonlinearities of Einstein's equations could be derived \cite{Rosen1}.  Einstein did not like such ideas when Rosen proposed them \cite{EinsteinvsRosen}.\footnote{I thank Dennis Lehmkuhl of the Einstein Papers Project at Caltech for bringing this correspondence to my attention.}  Neither did he like it when his assistant Robert Kraichnan was in the process of executing such a derivation \cite[p. xiv]{Feynman}.  (Bryce DeWitt, who transcended the GR \emph{vs.} particle physics divide to an unusual degree, did like Kraichnan's ideas \cite[xiv]{Feynman} \cite{DeWittDissertation}.)

One might see Einstein as predicting failure for such projects,  if success would be arriving at the 1915 field equations without having to know them already:  ``\ldots it would be practically impossible for anybody to hit on the gravitational equations'' without the ``exceedingly strong restrictions on the theoretical possibilities'' imposed by ``the principle of general relativity'' \cite{EinsteinGen}.  Is the principle of general relativity really needed?  As will appear below, (substantive) general covariance does play a key role, albeit as a \emph{lemma} rather than a premise, but generalized relativity of motion does not play any role\footnote{The concept of general covariance has become problematic especially in the last decade \cite{FriedmanJones,GiuliniAbsolute,PooleyGeneralCovariance,BelotGC}. For present purposes I am ignoring that problem. Anderson thought that this criterion was
equivalent to his preferred criterion of the absence of geometric objects that are the same (up to gauge equivalence) in all models \cite[pp. 88, 89]{Anderson}, but it isn't.}.  Such a derivation of Einstein's equations was successfully carried out in the 1950s-70s at  the classical level.  What this derivation implies about space-time and gravity is less clear, however.  Feynman has some remarks that could be construed as conventionalist \cite[pp. 112, 113]{Feynman}, shrinking the gulf between the initial flat geometry and the final effective curved geometry because the most convenient geometry can shift more easily than can the true geometry.  A complete story ought to take into account a notion of causality suitable for quantum gravity \cite[sect. 3.3.2]{ButterfieldIsham} \cite{NullCones,NullCones1}, but such issues will not be considered here.  There is no hint that substantive general covariance  is fed in and therefore easily recovered as has been claimed \cite{PadmanabhanMyth}; \emph{formal} general covariance is (inessentially) assumed, while \emph{substantive} general covariance is concluded, a quite different notion \cite{BergmannLectures,Anderson,StachelGC}.

The task of this paper is to show that this derivation of Einstein's equations, though quite compelling, is partly less novel  and is not  ontologically laden with flat space-time geometry than it has seemed.  Uncovering areas of overlap between general relativist and particle physics views might lead to further rapprochement.  A previous paper invoked the particle physics tradition as a foil for Einstein's 1913-15 physical strategy, which used somewhat similar ingredients including a key role for conservation laws and an analogy to Maxwell's electromagnetism \cite{EinsteinEnergyStability}.  There the thrust was how much light particle physics sheds on the processes of discovery and justification for Einstein's equations.  Here the direction of benefit is partly reversed:  particle physicists could have arrived at Einstein's equations much earlier if they had made use of Noether's converse Hilbertian assertion.  The derivation is also seen to be more ecumenical than one would have expected. Both of these works are thus early efforts at relating particle physics and the history of General Relativity.


\section{Noether's Converse ``Hilbertian Assertion'' about `Improper' Conservation Laws}

While Noether's work did not quickly get the explicit recognition that it deserved beyond initial endorsement by Klein \cite{KosmannSchwarzbachNoether}, by now it is deservedly a standard topic in the philosophy of physics \cite{BradingBook}.  
Besides Noether's first theorem deriving conserved currents from rigid symmetries (finitely many parameters) of the Lagrangian, and her second theorem deriving identities among the Euler-Lagrange equations, there are additional results of interest in Noether's paper.  These include, among other things  \cite{PetrovNoether}, many often neglected converse results, a proof of Hilbert's claim about the `improper' form of general relativistic energy-momentum conservation (the Hilbertian assertion), and, crucially for present purposes, a proof of the converse Hilbertian assertion:  improper conservation laws imply a (substantively) generally covariant action.

Noether's converses results seem to attract little attention even in works devoted to Noether's theorems.  There is an older literature that paid some attention to converses  \cite{FletcherConservation,DassNoether2,BoyerNoether,PalmieriConverseNoetherParticle,CandottiConverseNoetherField,CandottiNoetherField,CandottiConverseNoetherAJP,RosenNoetherField,CarinenaConverseNoether,FerrarioPasseriniNoether}.  
The fact that such works are most readily found before Noether's paper became widely available again (in terms of physical copies and language) due to a published English translation in 1971 \cite{Noether} is likely no coincidence.  Such works might well be based on second-hand reports \cite{HillNoether} that did not capture the full content of Noether's paper \cite{OlverLie}, including Noether's own emphasis on the converses.  Accord to Peter Olver, after 1922 
\begin{quote}
the next significant reference to Noether's paper is in a review article by the physicist Hill, [\cite{HillNoether}], in which the special case of Noether's theorem discussed in this chapter was presented, with implications that this was all Noether had actually proved on the subject.  Unfortunately, the next twenty years saw a succession of innumerable papers either re-deriving the basic Noether theorem 4.29 or purporting to generalize it, while in reality only reproving Noether's original result or special cases thereof.  The mathematical physics literature to this day abounds with such papers, and it would be senseless to list them here.   \cite[p. 282]{OlverLie}  \end{quote}

Among modern and philosophical works, there are apparently few that emphasize converses.  One is Katherine Brading's dissertation \cite[pp. 70-74]{BradingDissertation}, which emphasizes Noether's proof of the converse of her first theorem and uses it to undermine other authors' privileging of symmetries over conservation laws.  Likewise Harvey Brown and Peter Holland  doubt  that symmetries explain conservation laws, emphasize the converse, and include counterexamples to ideas that one might loosely have associated with Noether's first theorem  \cite{BrownHollandNoether1}.

If others have neglected Noether's converses, at least she thought them of special importance: 
\begin{quote} 
her own intent in writing her article had been ``to state in a rigorous fashion the significance of the principle and, above all, to
state the converse \ldots'' \cite[p. 52]{KosmannSchwarzbachNoether}. \end{quote} 
So she wrote in a referee report on a paper that covered similar ground to her own paper's but did not prove converses.

 The converse Hilbertian assertion is perhaps the most neglected of all; it is difficult to recall any attention being paid to it at all.  Perhaps it has seemed to be mathematical act of supererogation that would not benefit the working physicist. 
 What reason, after all, could one have for believing in improper conservation laws without already believing in Einstein's equations?   
Even Kosmann-Schwarzbach's book's discussion of the Hilbertian assertion is not very interested in the converse Hilbertian assertion (pp. 63, 64).  Emphasis is placed rather on the rigor added to what Hilbert had conjectured and the improper nature of the conservation laws as disanalogous to those following from rigid symmetries along the lines of Noether's first theorem (with antecedents in Lagrange and others \cite{KastrupNoetherKleinLie,EinsteinEnergyStability}).

Can one use the converse Hilbertian assertion to derive Einstein's equations?  To my knowledge deriving Einstein's equations \emph{via} improper conservation laws has never been attempted outside the particle physics tradition.  The logical equivalence of the gravitational field equations and the conservation laws has been noted  \cite{Anderson,EnergyGravity}, but that is still not enough. Such a derivation, to be sensible, would require some independent reason to believe in improper conservation laws.  Such independent reasons are not plentiful.  In the particle physics tradition such a derivation was  achieved 60 years ago \cite{Kraichnan}, but how it worked could use some clarification \cite{SliBimGRG}, especially to motivate the linear gauge freedom by avoiding ghosts \cite{PauliFierz,FierzPauli,NSSexlField,NSSexlLinear,VanN}. Even with that clarification  the connection to the Noether was not made.


\section{Particle Physics Spin $2$ Derivation(s)}

   Particle physicists have shown  Einstein's equations are what one naturally arrives for a local interacting massless spin-$2$ field, with good explanations for the detailed nonlinearity, general covariance, \emph{etc}. from nuts-and-bolts principles of (at least) Poincar\'{e}-covariant field theory. (Poincar\'{e} symmetry does not exclude a larger symmetry, as is especially clear from a Kleinian subtractive as opposed to Riemannian additive picture of geometry \cite{NortonKleinRiemann}.\footnote{The Kleinian picture assumes initially that coordinates are quantitatively meaningful for lengths, volumes, \emph{etc.} and then proceeds to strip them of meanings by larger symmetry groups. The Riemannian picture assumes that coordinates are quantitatively meaningless and then adds  structures  to define additional concepts such as volume, angle, length, \emph{etc.}}) 
 In 1939 it became possible to situate Einstein's theory \emph{vis-a-vis} the full range of relativistic wave equations and Lorentz group representations:  Pauli and Fierz recognized the equation for a massless spin 2 field as the source-free linearized Einstein equations \cite{Fierz,PauliFierz,FierzPauli}.  That same year Rosen wondered about deriving General Relativity's nonlinearities from an initially special relativistic starting point \cite{Rosen1}.  Work by Kraichnan, Gupta, Feynman, Weinberg, Deser \emph{et al.} eventually filled in the gaps, showing that, on pain of instability, Einstein's theory is basically the only option (eliminative induction, philosophers would say), with contributions by many authors, not all with the same intentions \cite{Weyl,Papapetrou,Gupta,Kraichnan,Thirring,Halpern,HalpernComp,Feynman,Wyss,OP,Weinberg65,NSSexlField,Deser,VanN,DeserQG,SliBimGRG,BoulangerEsole}.  One could also consider \emph{massive} spin 2 gravity if it works \cite{PauliFierz,FierzPauli,OP,FMS}, an issue apparently settled negatively in 1970-72 but reopened recently and now very active \cite{HinterbichlerRMP,deRhamLRR}. 
 To recall a punchy expression by Peter van Nieuwenhuizen, ``general relativity follows from special relativity by excluding ghosts''  \cite{VanN}. Even apart from the empirical fact of light bending (which van Nieuwenhuizen mentions) needed to refute scalar theories, the claim is slightly exaggerated \cite{SliBimGRG,MaheshwariIdentity,deRhamGabadadze,HassanRosenNonlinear}, but the point remains that it is difficult to avoid negative energy instability without strong resemblance to Einstein's equations.  Having negative-energy and positive-energy degrees of freedom interact seems likely to imply instability.  
And yet due to relativity's $-+++$ geometry, negative-energy degrees of freedom tend to crop up regularly if one is not careful \cite{FierzPauli,Wentzel,VanN}.
For a vector potential, if the spatial components have positive energy, then the temporal component will have negative energy unless one engineers it away, as occurs in Maxwell's theory.  Ghosts are almost always considered fatal (except in certain contexts where they are introduced as a technical tool).  What is to keep nothing from turning spontaneously into something and anti-something? What is possible soon becomes necessary in quantum mechanics; even energy conservation, supposed to exclude perpetual motion machines of the first kind, fails to stop the catastrophe.  Negative energy degrees of freedom were tacitly assumed not to exist in 19th century formulations of energy conservation, it would seem.  Lagrange considered whether positive energy was required for stability; he showed that bad things could happen if the potential were indefinite \cite{LagrangeMecaniqueAnalytiqueNouvelle}. 
 It did not occur to him to entertain negative \emph{kinetic} energy, however, something hardly conceivable apart from relativity and the de-materialization of matter into fields in the 20th century.  Lagrange showed that positive energy was stable with a separable potential  \cite{LagrangeMecaniqueAnalytiqueNouvelle}; the separability requirement was removed by Dirichlet \cite{Dirichletstability,Morrison}.
 According to Boulanger and Esole, 
\begin{quote} [i]t is well appreciated that general relativity is the unique way to consistently deform the
Pauli-Fierz action $\int \mathcal{L}_2$ for a free massless spin-2 field under the assumption of locality,
Poincar\'{e} invariance, preservation of the number of gauge symmetries and the number
of derivatives in $\mathcal{L}_2$ [references suppressed] \cite{BoulangerEsole}. \end{quote}

Derivations based on the canonical energy-momentum tensor or some relative thereof \cite{Gupta} perhaps do not savor as  strongly of flat space-time as do derivations with a flat metric \emph{tensor}. This impression might be a mistake due to failure to notice the extra gauge group that makes the flat metric less significant than the tensor notion suggests \cite{Grishchuk}. 
 The canonical tensor's advantage (if such it is) of not savoring as strongly of flat space-time is to some degree offset  by the gruelingly explicit character of the derivation.  This explicitness not only makes it harder to arrive at Einstein's theory (the massless case), but  also makes it difficult to generalize the results to include a graviton rest mass(es).  When efforts have been made to add a mass term using a canonical-based Belinfante energy-momentum, the result has been considered unique \cite{FreundNambu,FMS}.  By contrast  the Rosenfeld metric stress-energy approach used by Kraichnan has been generalized to show tremendous flexibility. Instead of a unique result for the tensor and scalar cases, the tensor case has grown to two one-parameter families \cite{MassiveGravity1}, to four one-parameter families \cite{MassiveGravity2}, to an arbitrary mass term with practically any algebraic self-interaction  \cite{MassiveGravity3}.  The scalar case likewise has been generalized from one case to a one-parameter family.  This one-parameter family is analogous to a tensorial 2-parameter family because the covariance-contravariance parameter (a continuum with  perhaps one hole \cite{OP}) and the density weight parameter cover the same ground; a $1 \times 1$ matrix is its own determinant.  Doubtless the scalar case could be generalized analogously to (\cite{MassiveGravity3}) to admit an arbitrary series (from cubic order onward) of mass plus self-interaction terms.

One can easily read spin $2$ derivations of Einstein's equations \cite{Gupta,Kraichnan,Feynman,Deser} without having a clear idea how they work.  When a canonical definition of stress-energy is used (improved by Belinfante symmetrization) as in Gupta's work, one can get lost in the explicit details, missing the forest for the trees.  When Rosenfeld's flat metric definition is used, the effect is more abstract and magical, along with the greater urge to wonder what the flat geometry means. One can clarify Kraichnan's derivation \cite{SliBimGRG}, partly by  drawing on Deser's use of a variational principle for free fields \cite{Deser} and van Nieuwenhuizen's ghost elimination \cite{VanN}.    Kraichnan used a bimetric change of variables from a gravitational potential and flat metric to an effective curved metric (the sum of the flat metric and curved metric) and flat metric, and showed that the flat metric disappears from the field equations.  \emph{The split of the  stress-energy tensor into a curl and a piece vanishing on-shell to derive Einstein's equations is a key step} \cite{SliBimGRG}. This fact is what allows one to recognize in effect Noether's converse Hilbertian assertion.  
 A connection to Noether's converse Hilbertian assertion has had to wait, however, to my knowledge.

Given this link to Noether's theorem and the Belinfante-Rosenfeld relation between canonical and metrical stress-energy, one could envisage a parallel derivation of Einstein's theory without the flat metric tensor, albeit much less convenient.  It is not coincidental that universal coupling derivations for massive scalar gravity using the canonical tensor have led only to a single theory \cite{FreundNambu}, because one needs to use the freedom to add a curl to the canonical tensor to accommodate second derivatives except in one case \cite{PittsScalar}. The ideal might be to use the  Belinfante-Rosenfeld equivalence the identity  to permit using the  Rosenfeld definition for calculations and  the Belinfante tensor (or it plus terms vanishing on-shell) for conceptual purposes.


\section{Aristotle on Inferring from Improper Conservation Laws to Einstein's Equations} 

One can take the leisurely pace of the linkage between Noether's proof of the converse Hilbertian assertion and the spin $2$ derivations as an indicator of the depth of the tragic split between general relativists and particle physicists, which this current paper aims to reduce somewhat.  Hilbert was insightful in taking improper conservation laws as characteristic of General Relativity, a comment that might be taken to suggest a derivation.  Noether's proof of the converse Hilbertian assertion could be taken providing the core of a derivation of Einstein's equations.  There is, however, an obvious problem, of a sort  discussed by Aristotle in the \emph{Posterior Analytics}. 
\begin{quote} 
The remainder of \emph{Posterior Analytics} I is largely concerned with two tasks: spelling out the nature of demonstration and demonstrative science and answering an important challenge to its very possibility. Aristotle first tells us that a demonstration is a deduction in which the premises are:

\begin{enumerate}  
  \item     true
  \item     {\bf primary} (\emph{prota})
  \item     {\bf immediate} (\emph{amesa}, ``without a middle'')
  \item     {\bf better known} or {\bf more familiar} (\emph{gn\^{o}rim\^{o}ôtera}) than the conclusion
  \item     {\bf prior} to the conclusion
  \item     {\bf  causes} (\emph{aitia}) of the conclusion
  \end{enumerate} %

\ldots 
The fourth condition shows that the knower of a demonstration must be in some better epistemic condition towards [the premises]\ldots. \cite[emphasis in the original]{AristotleLogicStanford}  
%
\end{quote} 
This fragment of Aristotle's theory of demonstration has an insight that one would presumably wish to retain:  if the premises are initially less plausible than the conclusion, then the argument is not very good.

In 1918 General Relativity was probably not known, but it was certainly seriously entertained.  Improper conservation laws were entertained only as a consequence of General Relativity.  Hence there was little prospect for regarding improper conservation laws as better known than or prior to Einstein's equations.  Particle physics changed that situation, partly by supplying a taxonomy in which one could fit General Relativity (massless spin $0$) and easily conceive a rival theory (massive spin $2$, at least \emph{prima facie}), thus making General Relativity less well known than it must have seemed after the 1919 bending of light success.  Shouldn't one be open to rival theories that made the same prediction?  Evidently some people were \cite{BrushLightBending}.  Particle physics also systematically implemented positive energy (avoiding ghosts) as a criterion of theory construction and theory choice \cite{PauliFierz,VanN}
and showed how a massless spin $2$ field satisfying the linearized Einstein equations is both a very natural path and one of very few paths to avoid ghosts with a symmetric rank $2$ tensor potential.  Indeed other plausible paths (unimodular/traceless massless spin 2/GR  \cite{UnruhUGR}, GR plus something like a scalar field \cite{SliBimGRG,TransverseFP}, and massive gravity \cite{MaheshwariIdentity,deRhamGabadadze}) are quite close to Einstein's equations.  In massive spin 2 gravity, Einstein's equations had at least one \emph{a priori} and empirically plausible rival \cite{OP,FMS}. Unless stability can be achieved in some other way, the positive energy requirement seems non-negotiable and, in turn, makes the linearization of Einstein's equations difficult to avoid, apart perhaps from a graviton mass term. The principle of universal coupling, which is another key part of the particle physics derivations, in fact was a part of Einstein's 1913-15 physical strategy as expressed in the \emph{Entwurf} with Grossmann:
\begin{quote} 
These equations satisfy a requirement that, in our opinion, must be imposed on a relativity  theory of gravitation; that is to say, they show that the tensor $\theta_{\mu\nu}$  of the gravitational field acts as a field generator in the same way as the tensor $\Theta_{\mu\nu}$ of the material processes.  An exceptional position of gravitational energy in comparison with all other kinds of energies would lead to untenable consequences. \cite{EinsteinEntwurf}  \end{quote} 

While ghost avoidance leaves nonlinearities quite unspecified as long as they aren't too strong, universal coupling provides a close link between the linearized  and exact nonlinear Einstein equations.  Improper  conservation laws follow \cite{SliBimGRG}: the stress-energy tensor is a piece vanishing on-shell and a curl. 
Thus it became possible to regard improper conservation laws as better known than Einstein's equations. Almost as van Nieuwenhuizen said, ``general relativity follows from special relativity by excluding ghosts'' \cite{VanN}.
  Aristotle's priority clause can be satisfied at least in the counterfactual history of how physics presumably would have progressed without Einstein \cite{Feynman,OhanianEinsteinMistakes}. It isn't clear why science should be forever held captive to historical accidents, so that is good enough.  Hence particle physics makes it plausible to argue for Einstein's equations using Noether's converse Hilbertian assumption, positive energy, \emph{etc.}

What about the principle(s) of equivalence?  Sometimes universal coupling is associated with the strong equivalence principle \cite[p. xiv]{Feynman}.
But universal coupling also leads to massive gravities \cite{FMS,MassiveGravity1,MassiveGravity2,MassiveGravity3}, for which there is a clear distinction between gravity and inertia. The identity of gravity and inertia is another strong  meaning sometimes associated with the principle of equivalence  \cite[pp. 37-39, 81]{FriedmanDynamicsReason}.  Avoiding multiply ambiguous terms like the principle of equivalence, one can safely say that the identity of gravity and inertia is not assumed in the particle physics derivation (massless or massive), and that such identity is clearly false at the end of the derivation in the massive case.


\section{(No) Ontology of Particle Physics Derivations} 

It is not terribly obvious what the ontology suggested by the spin-$2$ derivations of Einstein's equations is.   One often enough reads that the derivation shows that Einstein's theory is rendered just another field theory in Minkowski space-time, within special relativity,  or similar expressions. Such conclusions are especially tempting if one uses a flat metric \emph{tensor} (in the sense of having a nontrivial transformation rule under general coordinate transformations), not simply a matrix $diag(-1,1,1,1).$  Eventually it is concluded that the flat background metric is ``unobservable,'' which usually is supposed to mean or at least to imply that it doesn't really exist, perhaps \cite{Thirring}.  But how does coalescence of the flat geometry and the gravitational potential make the flat geometry cease to exist? 
One key issue pertains to whether one thinks in terms of Riemannian additive or Kleinian subtractive pictures of geometry \cite{NortonKleinRiemann}.  General relativists, historians and philosophers since the late 1970s have tended to default to a Riemannian additive picture, according to which Special Relativity is about an enormously impressive crystalline object, Minkowski Space-Time (or even Spacetime), which controls everything.  Particle physicists include toward a Kleinian subtractive picture, so that relativity is rather about having the Poincar\'{e} covariance group, which is certainly compatible with having an even larger covariance group.  Among philosophers,  calling Minkowski space-time a ``glorious nonentity'' is reminiscent of Klein's subtractive strategy \cite{BrownPooleyNonentity}.  Both Riemannian additive and Kleinian subtractive strategies are sometimes useful and illuminating.  

While some of the mathematics of the spin-$2$ derivations of Einstein's equations has a special relativistic feel---consider the title of Kraichnan's classic paper ``Special-Relativistic Derivation of Generally Covariant Gravitation Theory''---an additional gauge group emerges.  The emergence of an additional gauge group deprives of physical meaning the precise quantitative relationship between the effective curved metric and the/a flat background metric, making the flat space-time(s) elusive \cite{Grishchuk,NortonConvention,BrazilLocalize}, a point made early in little-known work by William Band \cite{Band1,Band2} and conceded but still insufficiently attended by Nathan Rosen in his application to gravitational energy localization \cite{RosenAnn}.  \emph{Which} flat metric underlies the effective curved geometry?  None in particular, one might say.  It is plausible that such a neo-traditionalist ontology is confusing to many, is (or would have been) attractive to some such as Lotze \cite[pp. 248, 249]{Lotze} (see also \cite[pp. 288, 299, 408]{TorrettiPhilosophy}), \emph{perhaps} Poincar\'{e} \cite{Poincare}, and Logunov \cite{LogunovBook}, and unattractive to others.  But the fact remains that the derivation of Einstein's \emph{equations} is quite compelling, much better than the competition involving Principles.  If one aspires to take the flat background seriously, then causality, ironically, requires reintroducing gauge freedom \cite{MassiveGravity1} (partly akin to privileging the Stueckelberg form of massive electromagnetism with gauge compensation fields \cite{Ruegg} over the primordial non-gauge form, and then restricting the gauge freedom with inequalities).  Hence some of the innovative features of General Relativity are obligatory anyway.

One could reduce the ontological confusion by getting rid of the flat metric tensor (or should one say, tensors, in light of the extra gauge group) in favor of the numerical matrix $diag(-1,1,1,1),$ such as one finds in some of this work already \cite{Gupta,OP}.   Much of the value of using a flat metric tensor(s) pertains to defining the stress-energy tensor.  The Rosenfeld stress-energy tensor is defined using a flat background metric, but one momentarily relaxes flatness to take a variational derivative, and then restores flatness again \cite{RosenfeldStress,Kraichnan,Anderson,Deser,GotayMarsden,SliBimGRG}, getting much of the benefit of Hilbert's definition without requiring the existence of the gravitational field. This definition is enormously convenient, but can cause confusion, both mathematical and conceptual.   That the Rosenfeld stress-energy tensor is just a trick has been urged in the context of Deser's spin-$2$ derivation: 
\begin{quote} 
\ldots $T_{\mu\nu}$ is the stress-tensor of the linear action of
equation (4). It is very simply computed in the usual (Rosenfeld) way as
the variational derivative of $I^L$ with respect to an auxiliary contravariant
metric density $\psi^{\mu\nu}$, upon writing $I^L$ in `generally covariant form', $I^L(\eta \rightarrow \psi),$
with respect to this metric. Note that this does not presuppose any geometrical
notions, being merely a mathematical shortcut in finding the
symmetric stress-tensor of $I^L$.  We could also obtain it by the (equivalent)
(Belinfante) prescription of introducing local Lorentz transformations.
\cite{Deser} \end{quote}

 By the Belinfante-Rosenfeld equivalence theorem, one can replace the Rosenfeld stress-energy tensor with the canonical tensor + certain terms with identically vanishing divergence + certain terms proportional to some field equations.  Actually calculating the amended canonical tensor might be unpleasant, but one could use the Belinfante-Rosenfeld equivalence theorem in the other direction so that the Rosenfeld stress-energy tensor is just a \emph{calculating trick}.  All conceptual work is done by the Belinfante modified canonical tensor (or it plus terms vanishing on-shell), but the calculations are done using Rosenfeld's trick.  Thus one evades an objection \cite{PadmanabhanMyth} that has been made to spin-$2$ derivations of Einstein's equations; others have been addressed \cite{MassiveGravity1}. One could avoid the additional gauge group, the topological limitations implied in introducing a flat metric \cite{AshtekarGeroch}, and most ontological connotations of a flat geometry(s) by using the matrix $diag(-1,1,1,1)$ instead of a flat background metric. This signature matrix is also intimately involved in nonlinear realizations of the `group' (using the term loosely) of general coordinate transformations \cite{OP,PittsSpinor}, which permit spinors in coordinates without a tetrad.  While one cannot take a variational derivative with respect to this numerical matrix, one does not need to do so, using the Belinfante-Rosenfeld equivalence as described above.  Conceptual matters are handled using the modified canonical tensor, while calculations can be handled with Rosenfeld's trick.  To implement this project, one can insert the Belinfante-Rosenfeld equivalence theorem into some suitable spin-$2$ derivation that uses the Rosenfeld tensor, 
preferably a version sufficiently flexible to display the previously unrecognized full generality of spin-$2$ derivations in cases where the graviton mass is not $0$ \cite{MassiveGravity3}.  Thus one overcomes a traditional weakness in spin-$2$ derivations using the canonical tensor, namely, arriving at only one theory and even that with much gritty detailed calculation \cite{FreundNambu,FMS}; such generality would otherwise likely appear fiendishly difficult and/or unnatural.  The canonical tensor and modifications thereof have the additional virtues (compared to the Rosenfeld metric stress-energy tensor) of being directly related to the translation symmetries that induces conservation and of being nontrivial (at least off-shell) even in topological (metric-free) field theories \cite{Burgess}.


\section{Conclusion}

It appears, then, that the particle physics derivation of Einstein's equations  should be quite attractive to everyone.  It involves principles that one could hardly avoid on pain of explosive instability.  It avoids  principles (including generalized relativity of motion or the identity of gravity and inertia) that could easily be false, though there is no harm if one finds them plausible. On the other hand the derivation is surprisingly innocent metaphysically, involving no ontological commitment to a flat space-time metric tensor.  Perhaps it should be more generally embraced, even by general relativists, rather than viewed as the special property of particle physicists.  Given that a core idea in the particle physics derivation is Noether's converse Hilbertian assertion, general relativists already do have  priority on part of the derivation.  This surprising convergence might set an example for additional fruitful work overcoming the general relativist \emph{vs.} particle physicist divide.


\section{Appendix}

The way that the split of the conserved stress-energy complex into a piece proportional to the gravitational field equations and a piece with automatically vanishing divergence plays a role in the spin $2$ derivation of Einstein's equations is worth recalling.  This derivation \cite{SliBimGRG} is based largely on that of Kraichnan \cite{Kraichnan} \cite[pp. xiii, xiv]{Feynman} but with a number of improvements.  The basic variables in this approach are the gravitational potential $\gamma_{\mu\nu}$ and the flat metric $\eta_{\mu\nu}$, along with (bosonic) matter fields $u$ that can be any kind of geometric object fields, though indices are suppressed. Gravity is assumed to have some free field equations derived from a (presumably quadratic) action $S_f[\gamma_{\mu\nu}, \eta_{\mu\nu}]$. 

In an effort to avoid negative-energy degrees of freedom, one can  require that $S_{f}$ change only by a boundary term under the infinitesimal gauge transformation \begin{eqnarray}  \gamma_{\mu\nu} \rightarrow  \gamma_{\mu\nu}  +  \partial_{\mu} \xi_{\nu} + \partial_{\nu} \xi_{\mu},  \label{gaugeinv} \end{eqnarray} $\xi_{\nu}$ being an arbitrary covector field. $\partial_{\mu}$ is the flat covariant derivative built from the flat metric $\eta_{\mu\nu}.$ 
(This condition is a bit stronger than necessary if the action does not imply higher derivatives in the field equations, but can be too weak if there are higher derivatives.  It is a good place to start, however.) 
  In the special case that the Lagrangian density is a linear combination of terms quadratic in first derivatives of the $\gamma_{\mu\nu}$, and free of algebraic and higher-derivative dependence on $\gamma_{\mu\nu}$, the requirement of gauge invariance uniquely fixes coefficients of the terms in the free field action up to a boundary term, giving linearized vacuum general relativity \cite{Ohanian,Hakim}.
       For any $S_{f}$ invariant in this sense under (\ref{gaugeinv}), the free field equation is identically divergenceless:   \begin{eqnarray} \partial_{\mu} \frac{\delta S_{f}  } {  \delta \gamma_{\mu\nu} }  = 0. \label{gaugeresult} \end{eqnarray}  This is the free field generalized Bianchi identity.  

As Einstein said in 1913, if a source is introduced, it is reasonable to have all stress-energy, for both gravity and matter fields $u$, serve as a source in the same way \cite{EinsteinEntwurf}.  Using the Rosenfeld stress-energy tensor from varying the (unknown) full action $S$ with respect to $\eta_{\mu\nu}$ (with $\gamma_{\mu\nu}$ and $u$ constant), one seeks the field equations from the (unknown) full action $S$ for gravity:
 \begin{eqnarray} \frac{\delta S}{\delta \gamma_{\mu\nu} } = \frac{\delta S_{f} }{\delta \gamma_{\mu\nu} } - \lambda \frac{\delta S}{\delta \eta_{\mu\nu} },  \label{Universal} \end{eqnarray}   where it turns out eventually that  $\lambda = - \sqrt{32 \pi G}$. 
This stress-energy tensor includes gravitational stress-energy; at this stage one doesn't know that an extra gauge group emerges and that the gravitational 
energy-momentum has the peculiar properties that it has in General Relativity.  (In a non-Rosenfeld form of such a derivation, one would not yet know that the gravitational energy-momentum is only a pseudo-tensor rather than a tensor. One could derive the linear analog of that fact if one takes the free field Lagrangian density to be quadratic in first derivatives of $\gamma_{\mu\nu},$ however \cite{Fierz}.)

One is free to make a change of variables in $S$ from $\gamma_{\mu\nu}$ and $\eta_{\mu\nu}$  to $g_{\mu\nu}$ and $\eta_{\mu\nu}$, where \begin{eqnarray} g_{\mu\nu} = \eta_{\mu\nu}  - \lambda \gamma_{\mu\nu}. \end{eqnarray}
 No assumption is made that $g_{\mu\nu}$ has chronogeometric significance; it emerges later as a result that $g_{\mu\nu}$ is the only quantity that could be an observable metric.  
  Equating coefficients of the variations gives \begin{eqnarray}  \frac{\delta S}{\delta \eta _{\mu\nu}} |\gamma =   \frac{\delta S}{\delta \eta_{\mu\nu}} |g  +  \frac{\delta S}{\delta g_{\mu\nu}}   \end{eqnarray} and  \begin{eqnarray}     \frac{\delta S}{\delta \gamma _{\mu\nu}}  =  - \lambda \frac{\delta S}{\delta g _{\mu\nu}}.  \label{ELVarChange} \end{eqnarray} 
Putting these two results together gives \begin{eqnarray}  \lambda \frac{\delta S}{\delta \eta_{\mu\nu}} |\gamma =  \lambda   \frac{\delta S}{\delta \eta_{\mu\nu}} |g  -  \frac{\delta S}{\delta \gamma_{\mu\nu}}.   \label{StressSplit} \end{eqnarray}  Equation (\ref{StressSplit}) splits the Rosenfeld stress-energy tensor into one piece that vanishes when gravity is on-shell and one piece that does not. That is one of the two types of terms that the converse Hilbertian assumptions allows.  
  Using this result in (\ref{Universal}) gives  \begin{eqnarray} \lambda \frac{\delta S}{\delta \eta _{\mu\nu}} |g =  \frac{\delta S_{f}}{\delta \gamma _{\mu\nu}}, \label{Key} \end{eqnarray} which says that the free field Euler-Lagrange derivative must equal (up to a constant factor) that part of the total stress tensor that does not vanish when the gravitational field equations hold.  Using the linearized Bianchi identity (\ref{gaugeresult}), 
 one derives  \begin{eqnarray} \partial_{\mu} \frac{\delta S}{\delta \eta _{\mu\nu}} |g = 0,  \label{curl} \end{eqnarray} which says that the part of the stress tensor not proportional to the gravitational field equations has identically vanishing divergence (on either index), \emph{i.e.}, is a (symmetric) ``curl'' \cite{Anderson}. This is the other type of term allowed by the converse Hilbertian assertion.  And there is nothing left of the stress-energy tensor:  those two terms, a piece proportional to the gravitational field equations and a piece with identically vanishing divergence, are the whole thing.  
The quantity $\frac{\delta S}{ \delta \eta _{\mu\nu}} |g$, being symmetrical and having identically vanishing divergence on either index, is of the form \begin{eqnarray}    \frac{\delta S}{ \delta \eta _{\mu\nu}} |g = \frac{1}{2}  \partial_{\rho} \partial_{\sigma} (  {\mathcal{M}} ^{[\mu\rho][\sigma\nu]} +   {\mathcal{M}} ^{[\nu\rho][\sigma\mu]} )  + b \sqrt{-\eta} \eta^{\mu\nu} \end{eqnarray} \cite{Wald} (pp. 89, 429) \cite{Kraichnan,SliBimGRG}, where ${\mathcal{M}} ^{\mu\rho\sigma\nu}$ is a tensor density of weight $1$ and $b$ is a constant.  This result follows from the converse of Poincar\'{e}'s lemma in Minkowski spacetime. One can gather all dependence on $\eta_{\mu\nu}$ (with $g_{\mu\nu}$ independent) into one term, writing  \begin{eqnarray}   S = S_{1} [g_{\mu\nu}, \cancel{\eta_{\mu\nu}}, u] + S_{2}[g_{\mu\nu}, \eta_{\mu\nu}, u]. \end{eqnarray} If    \begin{eqnarray}    S_{2} = \frac{1}{2} \int d^{4}x R_{\mu\nu\rho\sigma} (\eta) {\mathcal{M}} ^{\mu\nu\rho\sigma} (\eta_{\mu\nu}, g_{\mu\nu}, u ) + \int d^{4}x \alpha^{\mu},_{\mu} + 2 b \int d^{4}x \sqrt{-\eta},  \end{eqnarray} then $ \frac{\delta S_{2} }{  \delta \eta _{\mu\nu}}  |g $ has just the desired form, while $S_{2}$ does not affect the Euler-Lagrange equations \cite{Kraichnan,SliBimGRG}. ($\int d^{4}x \alpha^{\mu},_{\mu} $ is any boundary term that one likes.)  Thus the Euler-Lagrange equations arise entirely from $S = S_{1} [g_{\mu\nu}, \cancel{\eta_{\mu\nu}}, u]$:  the flat metric and the gravitational potential have merged, so the flat metric alone is unobservable and the only candidate for a metric is  $g_{\mu\nu}$.

Thus the stress-energy tensor turns out to be \emph{just} a term proportional to the Euler-Lagrangian equations and a term with identically vanishing divergence, and then one arrives at an action with Euler-Lagrange equations involving only a curved metric and matter fields, not a separate flat metric. 
Thus besides the (here trivial) formal general covariance, one has an additional gauge freedom to alter the flat metric tensor while leaving the  curved metric and matter fields alone, or, alternately, to alter the curved metric and matter fields by what looks like a coordinate transformation while leaving the flat metric alone  \cite{Grishchuk,NortonConvention,BrazilLocalize,SliBimGRG}.  If one fixes the coordinates to be Cartesian, then one has $\eta_{\mu\nu} = diag(-1,1,1,1)$ and the additional gauge freedom looks like a coordinate transformation in single-metric General Relativity (at least for transformations connected to the identity).  One thus de-Rosenfeldizes the result and arrives at what one usually considers a substantively generally covariant action $S_1[g_{\mu\nu}, u]$, as in the converse Hilbertian assertion.  Thus the spin $2$ derivation is, apart from the well-motivated and crucial elimination of ghosts, largely the  converse Hilbertian assertion all over again  with a glossy Rosenfeldized form using $\eta_{\mu\nu}$ and a {symmetric} curl term.   
  One can show that $\lambda = -\sqrt{32 \pi G}$ by, \emph{e.g.}, requiring proper normalization of the ($\Gamma-\Gamma$-like) bimetric General Relativity Lagrangian density of  Rosen \cite{Rosen1}.

Universal coupling turns out to deform the free field gauge invariance into a nonlinear gauge invariance.  (If one took a theory that had the linear gauge freedom but no nonlinear gauge freedom, it would have to violate universal coupling in order to escape the derivation above. A paper by Wald is also relevant \cite{WaldSpin}.) Chan and Fr{\o}nsdal provide a helpful summary.
 \begin{quote}  The apparently miraculous success of the original Gupta program has been convincingly explained by the analysis of Thirring [reference to \cite{Thirring,ThirringFdP}] and others. Namely, the structure of the full, nonlinear and non-Abelian gauge algebra of general relativity stands revealed upon completion of the first stage of Gupta's program. The geometric interpretation is immediate, and the full nonlinear action follows  from it. The essence of Gupta's method is to notice that the invariance of the free Lagrangian is equivalent to a degeneracy of the free wave operators (linearized Bianchi identities) and that this degeneracy leads to strong constraints on the form of interactions. This in turn implies the existence of a deformed invariance group of the perturbed (interacting) Lagrangian. \cite{FronsdalMass2} \end{quote} 
 That preservation of gauge symmetry is an important resource for avoiding ghosts at the nonlinear level.  It is possible to have nonlinear ghosts without linear ghosts  \cite{DeserMass}, a fact much discussed, and in some cases circumvented, in massive gravity \cite{deRhamGabadadze,HassanRosen}.   Another way to have nonlinear ghosts without linear ghosts would be to take, say, General Relativity, expand it perturbatively (which gives an infinite series except in the case of two fractionally weighted choices of fields \cite{DeWitt67b}), and then truncate the series at any finite order.  For all but the first few choices, such a theory will satisfy our empirical evidence for the weak principle of equivalence, but the theory will be bimetric   \cite{BlanchetNonmetric}. Thus it will have no gauge freedom and so will have six field degrees of freedom, not two as in GR.  One of them will be a ghost due  to the indefinite kinetic term of GR \cite{Wald} and any approximation thereof. Whereas a single-metric imagination renders one unable to conceive of such theories, particle physics enables one to conceive and refute them.  A few authors have suggested that one can live with ghosts in certain cases \cite{HawkingGhost}.  Perhaps so.  But that is a difficult road, especially under quantization where the threat of spontaneous production of arbitrarily many positive-and negative-energy quanta arises, so one needs to provide a detailed story for why such a theory is stable.  In short, mere empirical evidence is too weak to rule out infinitely many theories, almost empirically equivalent to GR, that likely are unstable due to nonlinear ghosts.  The no-ghost condition does important work to motivate Einstein's equations at both linear and distinctively again at nonlinear orders, work that mere empirical evidence can never do.  Again the superiority of the spin $2$ derivation over principles of equivalence and the like appears.  Norton has noted that Einstein's argumentation is often leaky and that one is well advised to seek eliminative inductions \cite{NortonEliminative}.  The spin $2$ derivation provides them.  

 This spin $2$ derivation admits some generalizations, such as covariant or contravariant metrics  of (nearly) arbitrary  density weight \cite{Kraichnan,Kraichnan2,MassiveGravity1}, an orthonormal (co)tetrad (of almost any density weight), thus facilitating spinors \cite{DeserSupergravity,MassiveGravity2}, and  nonlinear field redefinitions \cite{MassiveGravity3}.  While these derivations don't yield anything new for massless spin-$2$ gravity, they  yield an enormous variety for massive spin $2$ gravity if one requires only the kinetic term (the part with derivatives) to have gauge symmetry to try to avoid ghosts.  Whether the theories are viable requires additional investigation beyond these criteria because they might have ghosts, tachyons, a bad massless limit, or some other pathology. Massive theories, if viable, violate both the principle of generalized relativity and the identification of gravity with inertia.  One can also weaken the universal coupling condition to cover only the traceless part and leave an extra scalar (density) degree of freedom in the theory \cite{SliBimGRG}, thus arriving at scalar-tensor theories with the cosmological constant as a constant of integration.

One could also consider global topological issues. Flat metric tensors do make some demands on topology \cite{AshtekarGeroch}. But these demands are no stronger than spinors  require, because ``every flat space-time has spinor structure.'' \cite{GerochSpinor2}  


\section{Acknowledgements}
This work was supported by John Templeton Foundation grant \#38761.  Thanks to Harvey Brown for pointing me to references on the converse of Noether's first theorem, to  Dennis Lehmkuhl of the Einstein Papers project at Caltech for help with  the Einstein-Rosen correspondence, and to Alex Blum for help with Bryce Seligman DeWitt's dissertation.  Thanks also  to anonymous referees for helpful comments.  
%





\end{document}